# Comparison of P300 Responses in Auditory, Visual and Audiovisual Spatial Speller BCI Paradigms

M. Chang[1], N. Nishikawa[1], Z.R. Struzik[2,3], K. Mori[4], S. Makino[1], D. Mandic[5], and T.M. Rutkowski[1,2]

[1]Life Science Center of TARA, University of Tsukuba, Japan; [2]RIKEN Brain Science Institute, Wako-shi, Japan; [3]The University of Tokyo, Tokyo, Japan; [4]Research Institute of National Rehabilitation Center for Persons with Disabilities, Tokorozawa, Japan; [5]Imperial College London, London, UK

Correspondence: T.M. Rutkowski, Life Science Center of TARA, University of Tsukuba, 1-1-1 Tennodai, Tsukuba, Ibaraki, Japan. E-mail: tomek@tara.tsukuba.ac.jp

*Abstract.* The aim of this study is to provide a comprehensive test of three spatial speller settings, for the auditory, visual, and audiovisual paradigms. For rigour, the study is conducted with 16 BCI-naïve subjects in an experimental set-up based on five Japanese hiragana characters. Auditory P300 responses give encouragingly longer target vs. non-target latencies during the training phase, however, real-world online BCI experiments in the multimodal setting do not validate this potential advantage. Our case studies indicate that the auditory spatial unimodal paradigm needs further development in order to be a viable alternative to the established visual domain speller applications, as far as BCI-naïve subjects are concerned.

*Keywords:* EEG, P300, auditory evoked potentials, multimodal BCI

## 1. Introduction

Contemporary brain computer interface (BCI) paradigms rely mostly on unimodal approaches [Wolpaw and Wolpaw, 2012]. Recently proposed solutions enhance the existing paradigms by adding spatial stimuli variability [Halder et al., 2010; Schreuder et al., 2010; Muller-Putz et al., 2006; van der Waal et al., 2012] in order to augment the brain-computer interfacing comfort or to boost the information transfer rate (ITR) achieved by the users. We tested this with 16 BCI-naïve subjects, in a simple five Japanese hiragana (a i u e o) spatial speller task to compare the interfacing accuracy variability and the users' subjective comfort. In order to do so, we designed three spatial tasks with visual letters and synthetic vowel representations originating from the congruent spatial directions using a virtual sound panning technique. We compare the results based on unimodal and bimodal spelling experiments and draw conclusions for possible future research directions.

## 2. Material and Methods

In the experiments reported in this paper, 16 BCI-naïve subjects took part (mean age 21.81 with a standard deviation of 0.75). All the experiments were performed at the Life Science Center of TARA, University of Tsukuba, Japan. The online EEG BCI experiments were conducted in accordance with the WMA Declaration of Helsinki - Ethical Principles for Medical Research Involving Human Subjects. The subjects of the experiments received a monetary gratification. The 200 ms long spatial unimodal (visual or auditory) or bimodal (audiovisual) stimuli were presented from five distinct spatial locations. In the case of visual and audiovisual speller paradigms, large size hiragana characters were flashed one at a time on a big computer display positioned in front of the subject (as in a usual oddball based P300 visual speller). In the case of the auditory modality, we designed a vector based sound amplitude panning application, which positioned a virtual sound image at a spatial location congruent with a visual letter (positioned at $-45°$, $-22.5°$, $0°$, $22.5°$, and $45°$ in front of the head). Each subject first conducted a short psychophysical test with a button press response to confirm understanding of the experimental set-up in each modality. During the online BCI experiments, the EEG signals were captured with a 16 active electrodes EEG amplifier system, g.USBamp by g.tec. The electrodes were attached to the following head locations Cz, CPz, POz, Pz, P1, P2, C3, C4, O1, O2, T7, T8, P3, P4, F3, and F4, as in the 10/10 extended international system (see topographic plot in Fig. 1). The ground and reference electrodes were attached at FCz and the earlobe respectively. The recorded EEG signals were processed by the in-house enhanced BCI2000 application using a linear discrimination analysis (LDA) classifier with features drawn from 0-600 ms event related potential (ERP) intervals. The sampling frequency was set to 512 Hz, the high pass filter at 0.1 Hz, and the low pass filter at 100 Hz, with a power line interference notch filter set in the 48-52 Hz band. The inter-stimulus interval (ISI) was set to 500 ms and each stimuli length was 200 ms. The subjects were instructed to spell five random sequences of hiragana letters, which were presented visually, audibly or audiovisually in each session respectively. Each target was presented ten times in a single spelling trial and the averages of ten ERPs were later used for the classification in order to make the experiment easier for novices.

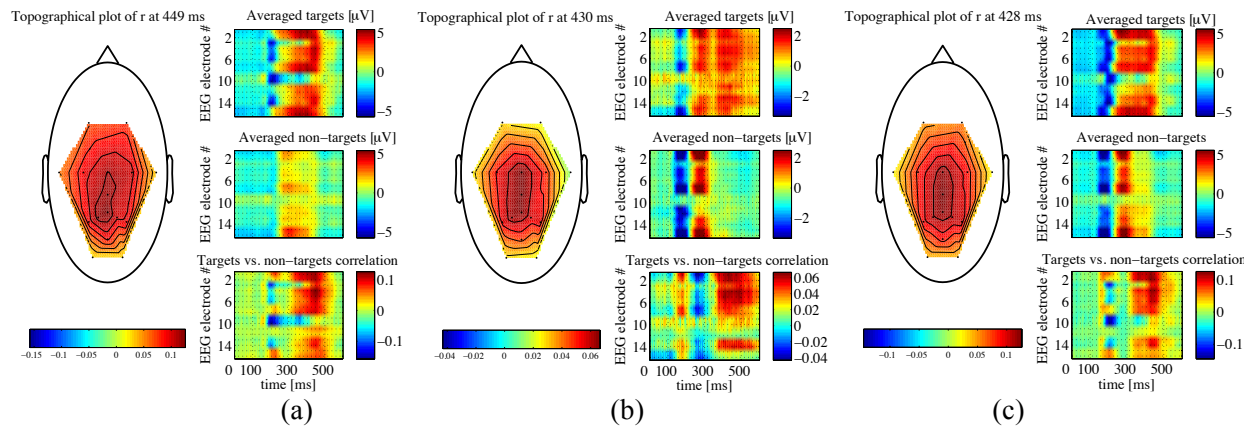

*Figure 1.* The averaged ERP responses of the 16 BCI-naïve subjects taking part in the experiment. Each panel (a), (b), and (c) presents topographical maps of target vs. non-target response correlation coefficient distributions at the most discriminative latency. The averaged ERPs are also presented separately, together with correlation coefficient time series on the right side of each panel. Panel (a) presents visual speller results; (b) auditory; and (c) audiovisual. The interesting "longer" P300 averaged response deflection visualized in the top panel (b), and in the correlation time series, was not reflected in better accuracy scores (see Table 1). The EEG electrode, represented by numbers on the vertical axes is shown in the right panels. The electrode order is as follows Cz, CPz, POz, Pz, P1, P2, C3, C4, O1, O2, T7, T8, P3, P4, F3, and F4.

## 3. Results and Discussion

The averaged ERP responses from the 16 BCI-naïve subjects are presented in Fig. 1 for the three tested modalities. The results shown in panel (b) of the Fig. 1 are encouraging, since a more distinct (longer latencies) P300 response is observed in the auditory unimodal paradigm. Unfortunately, this observation could not later be confirmed in the form of correct spelling accuracies in the online BCI experiments reported in Table 1, where the classical visual unimodal paradigm still resulted in the best scores. The topographic plots presented in Fig. 1 confirm the classical P300 response scalp positions with more localized distributions for auditory unimodal experiments. The preliminary, yet encouraging results presented call for more research on spatial auditory BCI paradigms in order better to utilize the encouraging "longer" P300 response.

*Table 1.* Summary of the online BCI interfacing results with the 16 naïve subjects in auditory, visual and audiovisual modalities.

| Modality | Mean accuracy | Standard deviation |
|---|---|---|
| Auditory | 52.5% | 27.2% |
| Audiovisual | 91.3% | 12.6% |
| Visual | 95.0% | 11.5% |

## Acknowledgements

This research was supported in part by the Strategic Information and Communications R&D Promotion Programme no. 121803027 of The Ministry of Internal Affairs and Communication in Japan, and by KAKENHI, the Japan Society for the Promotion of Science grant no. 12010738. We also acknowledge the technical support of YAMAHA Sound & IT Development Division in Hamamatsu, Japan.